\documentclass[11pt]{article}
\usepackage[a4paper]{geometry}
\usepackage[T1]{fontenc}
\usepackage{lmodern}
\usepackage{authblk}
\usepackage{amsmath,amssymb}
\usepackage{graphicx}
\usepackage{booktabs}
\usepackage{microtype}
\usepackage{listings}
\usepackage[numbers,sort&compress]{natbib}
\usepackage{xurl}
\usepackage[hidelinks]{hyperref}
\hypersetup{pdftitle={dexamine: A Python package for Uniswap event data on Ethereum},
  pdfauthor={Magnus Hansson}}
\title{\texttt{dexamine}: A Python package for Uniswap event data on Ethereum}
\author[1,2]{Magnus Hansson}
\affil[1]{Stockholm University}
\affil[2]{Swedish House of Finance}
\date{}

\begin{document}
\maketitle
\vspace{-2em}
\begin{center}
\small
\href{mailto:hansson.carl.magnus@gmail.com}{\nolinkurl{hansson.carl.magnus@gmail.com}}\\
ORCID: \url{https://orcid.org/0009-0004-4318-5145}\\
\end{center}

\begin{abstract}
\noindent
Decentralized exchanges record trading and liquidity provision on public blockchains,
but empirical analysis requires interpreting these records and linking them to
execution metadata. dexamine is a Python package that parses Uniswap v2 and v3 events
on Ethereum. It converts transaction receipt logs into observations of trades and
liquidity changes, with token quantities, pool state, transaction order, and gas
information. The package separates data retrieval, contract metadata, protocol
interpretation, and output construction. The repository provides recorded Ethereum
responses and an offline reproducible example, and version 1 has been used to
construct data for an empirical study of price discovery in decentralized markets.
\end{abstract}


\section{Introduction}
Decentralized exchanges allow users to trade digital assets through programs executed
on a blockchain. Uniswap v2 and v3 are automated market makers on Ethereum: trades
change the balances of liquidity pools, and prices follow rules implemented by their
contracts \cite{adams2020,adams2021}. Their public records provide data for empirical
studies of trading and liquidity provision. However, these records describe contract
execution and must be interpreted before they become economic observations.

Studies of decentralized exchange markets examine liquidity and trading costs
as well as prices \cite{lehar2025,barbon2026}. Research on arbitrage and transaction
ordering also shows why execution sequence and fees matter \cite{daian2020}.
A transaction's position within a block, execution fees, and the pool state following
a trade help describe the conditions under which it executed. One Ethereum transaction can
contain several trades or liquidity
events, each of which must remain distinguishable. A block timestamp alone does not
recover their execution order.

Ethereum exposes blocks, transactions, and event logs through JSON-RPC. Constructing
a dataset requires protocol-specific decoding, resolution of token metadata, conversion
of integer quantities into token units, and joins between events and execution
metadata. Repeating these transformations across projects creates opportunities for
inconsistent units, signs, and ordering. dexamine implements them in a reusable Python
library.
The researcher supplies transaction positions, defined by block number and transaction
index, together with a protocol selection and, optionally, a pool address, and receives records for the supported events.

Related tools address extraction and indexing at different levels. Ethereum ETL exports
blockchain records and selected token data \cite{ethereumetl}. cryo supports bulk
extraction into tabular formats and event decoding from supplied signatures
\cite{cryo2023}. Graph Node provides application-specific indexing and GraphQL queries
through subgraphs and can be operated locally \cite{graphnode}. Its retained information
depends on the chosen schema and mappings. TrueBlocks provides address-based transaction
indexing \cite{trueblocks}. These tools are discussed as related software; dexamine
does not import or invoke them. Researchers can use external tools to select
transactions before passing their positions to dexamine.

The contribution of dexamine is a common representation of Uniswap events joined to
their execution metadata. Generic extraction tools still require these transformations,
while a subgraph places them within an indexing service. A separate library allows
researchers to reuse RPC and contract-access facilities while incorporating the
protocol interpretation into their own pipelines. Transaction discovery, storage,
and statistical estimation remain independent stages.

Version 1 of dexamine has been used for
data construction in a study of price discovery in decentralized markets \cite{hansson2024}.
This application requires preserving the sequence of events and their associated
pool information. Reusing a parser provides consistent treatment of token units,
liquidity changes, and execution metadata across analyses.

The software, documentation, tests, and reproducible examples are available at
\url{https://github.com/HanssonMagnus/dexamine}.

\section{Software design and implementation}
\subsection{Installation and usage}
dexamine can be installed directly from the source repository:

\begin{lstlisting}[language=bash]
python -m pip install git+https://github.com/HanssonMagnus/dexamine.git@v1.1.0
\end{lstlisting}

The package requires Python 3.10 or newer, web3.py 6.15.0 or newer, and Requests
2.31.0 or newer. Live parsing requires access to an Ethereum JSON-RPC endpoint
retaining the requested historical blocks and transaction receipts and supporting
contract metadata calls. The recorded example in Section~\ref{sec:example} runs
offline without a node.

\subsection{Software architecture}
\begin{figure}[!ht]
\centering
\includegraphics[width=0.85\linewidth]{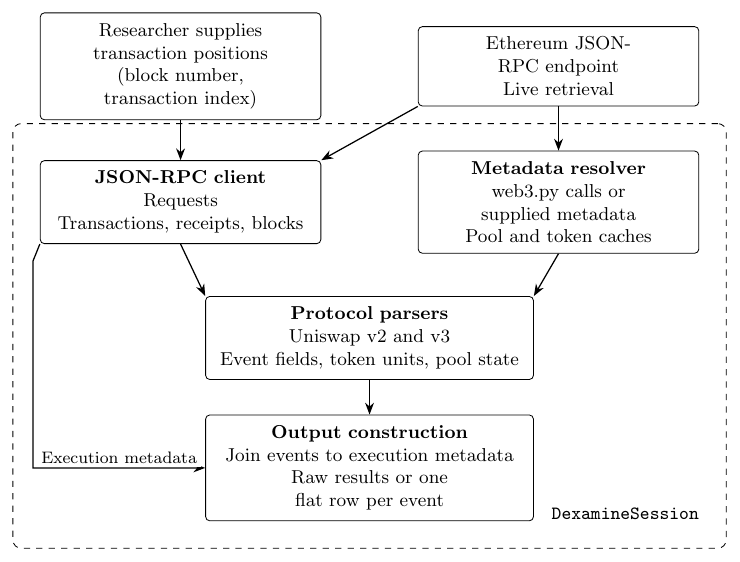}
\caption{Data flow within a session. The RPC client retrieves execution records;
the metadata resolver supplies pool and token information. Protocol parsers decode
receipt logs, and the output layer joins the resulting events to transaction and
block metadata. Offline replay supplies recorded RPC results and seeded metadata.}
\label{fig:architecture}
\end{figure}

Figure~\ref{fig:architecture} shows the separation of JSON-RPC retrieval, contract
metadata resolution, protocol parsers, and output construction. A reusable session loads contract interfaces
and retains pool and token metadata across calls. dexamine's JSON-RPC client uses
Requests \cite{requests} to retrieve transactions, receipts, and blocks over HTTP.
web3.py \cite{web3py} supplies contract calls for pool and token metadata and address
checksum conversion. Requests and web3.py are the two declared runtime dependencies.
The protocol parsers decode event topics and fixed-width data fields directly in
Python. The output layer joins these event records to execution metadata.

Requests for multiple positions are batched. Results are yielded incrementally, so
event records need not accumulate in memory. The metadata cache grows with the number
of distinct contracts encountered. Users can divide positions across processes with
one session per worker. Processing rates depend on endpoint latency and capacity,
batch size, and metadata cache misses; no throughput advantage over other extraction
tools is claimed here.

\subsection{Functionality}
The package supports Uniswap v2 and v3 swaps, mints, and burns on Ethereum mainnet.
Mints and burns describe additions and removals of liquidity. Event records include
token symbols and decimal precision, normalized token quantities, and event-specific
pool information. The sign conventions express liquidity additions as positive
quantities and removals as negative quantities; swap quantities describe the pool's
net token flows.

The parsers account for protocol differences. Uniswap v2 reports reserve updates in
a preceding \texttt{Sync} event. dexamine checks that the immediately preceding log
is a \texttt{Sync} from the same pool and skips an event if this condition fails.
For swaps, the reserve ratio supplies the pool's post-event marginal price before fees.
Uniswap v3 swap logs report price, tick, and active liquidity directly. The parser
uses these values to derive price and virtual reserves. These reserves describe the
local trading curve, rather than total pool balances; they are reported as missing
when active liquidity is zero.

The output can retain node payloads alongside parsed events or provide one flat row
per event. Flat records include block and transaction indices, the transaction hash,
and the event's zero-based position within the receipt's log list. This last field
is named \texttt{receipt\_log\_index}; it differs from Ethereum's block-wide
\texttt{logIndex}. Together, transaction identity and receipt position identify the
source event. Missing or inapplicable numerical fields are represented by
\texttt{None}, which becomes \texttt{null} in JSON.

Token amounts are scaled by their decimal precision. Normalized quantities use
floating-point arithmetic and can contain rounding error, so the output is intended
for empirical analysis rather than exact integer accounting. Retaining the original
logs allows users to recover integer quantities. Gas consumption and fee fields
apply to the entire transaction and are repeated across its events. They do not
allocate execution costs to individual trades.

Destination labels are constructed from the transaction's top-level destination.
Transactions with no destination are labeled \texttt{contract\_creation};
destinations matching a bundled list of Uniswap router addresses are
labeled \texttt{uniswap\_router}; and all remaining destinations are labeled
\texttt{other\_contract}. This classification does not reconstruct internal call
paths or establish trader identity, arbitrage, or maximal extractable value.

\subsection{Reproducibility and verification}
\label{subsec:verification}
Reproducibility depends on the software version, source responses, and resolved
metadata. Pool and token metadata are queried at the latest block and cached.
Changes to token metadata can therefore affect repeated runs even when historical
logs are unchanged. Retaining input responses and metadata supports replication.
The public \texttt{MetadataResolver.\allowbreak seed} method accepts recorded token and pool
metadata, normalizes addresses, and replaces the supplied cache entries. Missing
entries still trigger endpoint queries; callers retain responsibility for the
provenance of supplied metadata.
The endpoint must serve the requested historical blocks and receipts and support
metadata calls, but the parser does not require historical contract-state queries.
Users should also retain transaction hashes when selecting positions, since chain
reorganizations can change the transaction at a recent position.

Offline tests exercise event decoding, signed quantities, reserve ordering, pool
filtering, and missing values using recorded responses and constructed event logs.
An optional integration suite checks the public interface against a live endpoint.
Continuous integration runs tests, formatting, linting, and strict type checking on
Python 3.10 through 3.13. Installation instructions, field definitions, limitations,
and contribution guidance are supplied with the code.

\section{Illustrative example}
\label{sec:example}
Transaction 31 in Ethereum block 12,561,528, recorded on 3 June 2021, contains four
Uniswap v3 swaps across three pools. The following code parses all supported v3 events
in the transaction without restricting the pool address, then orders the rows by
their positions in the receipt:

\medskip
\begin{minipage}{\linewidth}
\begin{lstlisting}[language=Python]
from dexamine import DexamineSession

session = DexamineSession.from_node_url(node_url)
rows = session.parse_position(
    block_number=12561528,
    tx_index=31,
    protocol="uniswap_v3",
    exchange_pair_address=None,
    output_format="flat",
)
rows.sort(key=lambda row: row["receipt_log_index"])
\end{lstlisting}
\end{minipage}
\medskip

Table~\ref{tab:example} shows the four swaps in receipt order. The token symbols
follow each pool's token0/token1 ordering. Positive quantities enter the pool;
negative quantities leave it. USDC amounts are scaled by $10^6$, and DAI and WETH
amounts by $10^{18}$. The intervening logs include token transfers and other events
outside the v3 parser's supported set. Their presence explains the gaps between
receipt positions. The first and last swaps concern the same USDC/WETH pool, but
remain distinct observations within one transaction.

\begin{table}[!htbp]
\centering
\small
\begin{tabular}{rllrr}
\toprule
Receipt position & Token 0 & Token 1 & Amount 0 & Amount 1 \\
\midrule
3  & USDC & WETH & $-14008.253925$ & $5.000000$ \\
6  & DAI  & USDC & $-13983.724002$ & $14008.253925$ \\
10 & DAI  & WETH & $13983.724002$ & $-4.999395$ \\
13 & USDC & WETH & $-14003.702900$ & $4.999395$ \\
\bottomrule
\end{tabular}
\caption{All four Uniswap v3 swaps in the recorded transaction. Amounts are in token
units and rounded to six decimal places. Receipt positions are zero-based.}
\label{tab:example}
\end{table}

Each swap also reports the post-event price, tick, and active liquidity.
Table~\ref{tab:swap-state} shows the price and virtual reserves derived from that
state. Prices are expressed as token 0 per token 1: USDC per WETH, DAI per USDC,
and DAI per WETH for the respective pools. These are marginal prices after the
swap, rather than ratios of the exchanged amounts. Virtual reserves characterize
the local trading curve at the reported active liquidity; they are not the pool's
total token balances. In particular, the large DAI/USDC virtual reserves should not
be interpreted as assets held by that pool.

\begin{table}[!htbp]
\centering
\small
\begin{tabular}{@{}rrrrr@{}}
\toprule
Receipt position & Price & Tick & Virtual reserve 0 & Virtual reserve 1 \\
\midrule
3  & 2802.766431 & 196936  & 137336604.311 & 49000.374 \\
6  & 0.998748    & $-276312$ & 94086697387.618 & 94204630210.412 \\
10 & 2800.001801 & $-79378$  & 9064867.550 & 3237.451 \\
13 & 2802.194884 & 196938  & 137322600.609 & 49005.371 \\
\bottomrule
\end{tabular}
\caption{Post-swap pool state for the events in Table~\ref{tab:example}. Prices are
rounded to six decimal places and virtual reserves to three, in the corresponding
token units. Active liquidity and the encoded square-root price are retained in
the complete JSON output.}
\label{tab:swap-state}
\end{table}

The explicit sort recovers execution order even when different supported event
types are grouped by the parser. Pool and token addresses are retained in the
recorded inputs; symbols provide readable labels, not unique contract identifiers.
Gas usage of 476,588 and an effective gas price of 40 gwei apply to the whole
transaction and repeat on every row. Summing those fields across swaps would count
the same transaction cost four times. The \texttt{other\_contract} destination label
describes the top-level address match and does not establish trader identity or
the transaction's economic purpose.

The accompanying
\href{https://github.com/HanssonMagnus/dexamine/tree/d030990fa797c4b876bed2b7d474649bcfb1358e/paper/examples}{recorded example}
reproduces the output offline with dexamine 1.1.0. It contains the full transaction,
receipt, and block responses, the transaction hash and capture provenance, and
metadata for all three pools and tokens. The metadata was captured separately
through contract calls at the recorded block, rather than at the latest block.
The script supplies it through \texttt{MetadataResolver.\allowbreak seed} and
replays the public RPC client's \texttt{call} method while exercising the session,
parser, and output construction. Unexpected HTTP access is rejected.

The \verb|--check| option compares all four rows with the committed JSON;
\verb|--node-url| enables live retrieval with normal metadata resolution.
A separate verification script checks transaction and block identity, decodes the
archived metadata calls, and independently checks each swap's amounts and pool
state using ABI decoding and high-precision decimal arithmetic. Both checks run
in continuous integration without a node.

\section{Conclusions}
dexamine converts Uniswap v2 and v3 receipt logs into event observations linked to
Ethereum execution metadata. Its contribution is a reusable implementation of protocol
interpretation and data joins for empirical research. The library supports both
individual transactions and batched processing, with offline verification and an
executable example.

\section*{Acknowledgements}
I thank the TrueBlocks \cite{trueblocks} and Erigon \cite{erigon} teams for assistance with node operation
and indexing, and members of the Flashbots and Uniswap communities for discussions
of the protocols.

\section*{Code and data availability}
The source code for dexamine is available at
\url{https://github.com/HanssonMagnus/dexamine} under the GNU General Public License
v3.0 or later. This paper describes version
\href{https://github.com/HanssonMagnus/dexamine/tree/v1.1.0}{\texttt{v1.1.0}}.

The recorded Ethereum responses used in Section~\ref{sec:example} are included
under \path{paper/examples/data}. The example scripts, expected output, historical
metadata, and recorded inputs are distributed with the repository.

\bibliographystyle{unsrtnat}
\bibliography{paper}
\end{document}